\documentclass[article]{aamod}

\usepackage{natbib}
\bibpunct{(}{)}{;}{a}{}{,}
\usepackage{graphicx}
\usepackage{txfonts}

\begin{document}

\title{Multicomponent \ion{He}{i} 10830~{\AA} profiles in an active filament}

\author{C. Sasso\inst{1,2} \and A. Lagg\inst{1} \and S. K. Solanki\inst{1,3}}

\offprints{C. Sasso, \email{csasso@oacn.inaf.it}}

\institute{Max-Planck-Institut f\"ur Sonnensystemforschung,
Max-Planck-Str. 2, Katlenburg-Lindau, Germany \and INAF-Osservatorio
Astronomico di Capodimonte, Salita Moiariello 16, I-80131 Napoli,
Italy \and School of Space Research, Kyung Hee University, Yongin,
Gyeonggi 446-701, Korea}

\date{}

\abstract{}{We present new spectropolarimetric observations of the
chromospheric \ion{He}{i} 10830~{\AA} multiplet observed in a
filament during its phase of activity.}{The data were recorded with
the new Tenerife Infrared Polarimeter (TIP-II) at the German Vacuum
Tower Telescope (VTT) on 2005 May 18. We inverted the He Stokes
profiles using multiple atmospheric components.}{The observed He
Stokes profiles display a remarkably wide variety of shapes. Most of
the profiles show very broad Stokes $I$ absorptions and complex and
spatially variable Stokes $V$ signatures. The inversion of the
profiles shows evidence of different atmospheric
blue- and redshifted components of the \ion{He}{i} lines within the
resolution element ($\sim 1$~arcsec), with supersonic velocities
of up to $\sim 100$~km/s. Up to five different atmospheric components are
found in the same profile. We show that even these complex profiles
can be reliably inverted.}{}


\maketitle

\section{Introduction}

Filaments are typically elongated dark structures on the solar disk,
characterized by additional absorption in strong chromospheric
lines, like the \ion{H}{$\alpha$} line. They are the on-disc
counterparts of prominences seen above the solar limb and represent
dense and cool chromospheric gas suspended in the hot and thin
corona. The strength and structure of the magnetic field plays a key
role in keeping the prominence material from flowing down to normal
chromospheric levels. It is therefore of considerable interest to
determine the magnetic structure associated with a prominence. A
number of observations of the magnetic vector in prominences have
been carried out
\citep[e.g.,][]{leroy,casini,lin,Trujillo_nature,merenda,kuckein}.
All these investigations, except for the last one, have been restricted
to quiescent prominences. \\
In this paper we introduce spectropolarimetric observations in the
\ion{He}{i} 10830~{\AA} triplet of a filament located in an active
region that has been activated by a flare. Filaments also appear as
dark absorption features in the \ion{He}{i} 10830~{\AA} line. \\
Spectropolarimetry in the \ion{He}{i} triplet at 10830~{\AA} is an
important tool to determine the magnetic field vector in the solar
chromosphere \citep{Trujillo_nature,solanki_nature}. It is well
suited for the measurement of the magnetic vector near the base of
the solar corona because these lines are often narrow,
nearly optically thin, have a reasonable effective Land\'e factor
(see Table~\ref{tab:righe}) and are easily observed \citep{ruedi}.
Measurements of the polarization of the \ion{He}{i} 10830~{\AA}
radiation are also indicated as a useful new tool for the
diagnostics of the magnetic field in filaments
\citep{lin,Trujillo_nature}. Recently, \citet{kuckein} studied the
vector magnetic field of an active region filament by analyzing
spectropolarimetric data in the \ion{He}{i} 10830~{\AA} lines with
three different methods, one of these being, as in our case,
Milne--Eddington inversions. They find the highest field strengths
measured in filaments so far, around 600-700~G, and conclude that
strong transverse magnetic fields are present in active region
filaments. \\
The \ion{He}{i} 10830~{\AA} multiplet originates between the atomic
levels $2$ $^3S_1$ and $2$ $^3P_{2,1,0}$. It comprises a component
at 10829.0911~{\AA} with $J_u=0$ (hereafter referred to as Tr1),
and two components at 10830.2501~{\AA} with
$J_u=1$ (Tr2) and at 10830.3397~{\AA} with $J_u=2$ (Tr3) which are
blended at solar chromospheric temperatures. In the subsequent
discussion we call this blended absorption the
Tr2,3 component. \\
We concentrate on describing
the observations and the inversion of these profiles.
The He Stokes profiles in the activated filament reveal a remarkable
level of complexity, requiring up to five independent magnetic
components to reproduce. In a following paper the obtained maps of
the magnetic and velocity fields in the filament will be presented,
critically analyzed and interpreted in terms of prominence models.

\section{Observations}

\begin{table}
\caption{Spectral lines contained in the spectral window of the
TIP-II instrument (10825-10836~{\AA}).}\label{tab:righe}
\centering
\begin{tabular}{ccc}
  \hline\hline
  Line                       & Wavelength (\AA) & Effective Land\'e factor\\
  \hline
  \ion{Si}{i}                & 10827.09         & 1.5           \\
  \ion{He}{i} (Tr1)          & 10829.09         & 2.0           \\
  \ion{Ca}{i}                & 10829.27         & 1.00          \\
  \ion{He}{i} (Tr2)          & 10830.25         & 1.75          \\
  \ion{He}{i} (Tr3)          & 10830.34         & 1.25          \\
  \ion{H$_2$O} (telluric)    & 10832.11         &               \\
  \ion{Ca}{i}                & 10833.38         & 1.00          \\
  \ion{H$_2$O} (telluric)    & 10833.90         &               \\
  \ion{Na}{i}                & 10834.85         & 1.03          \\
  \hline
\end{tabular}
\end{table}
\begin{figure*}
\centering
\includegraphics[width=6.86cm,clip=true]{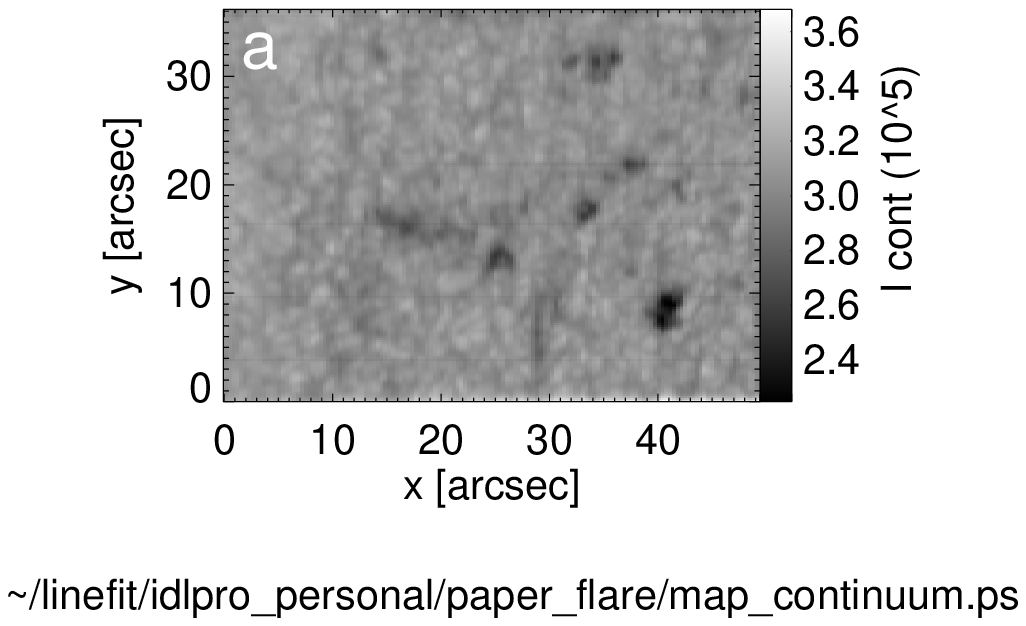}
\includegraphics[width=6.86cm,clip=true]{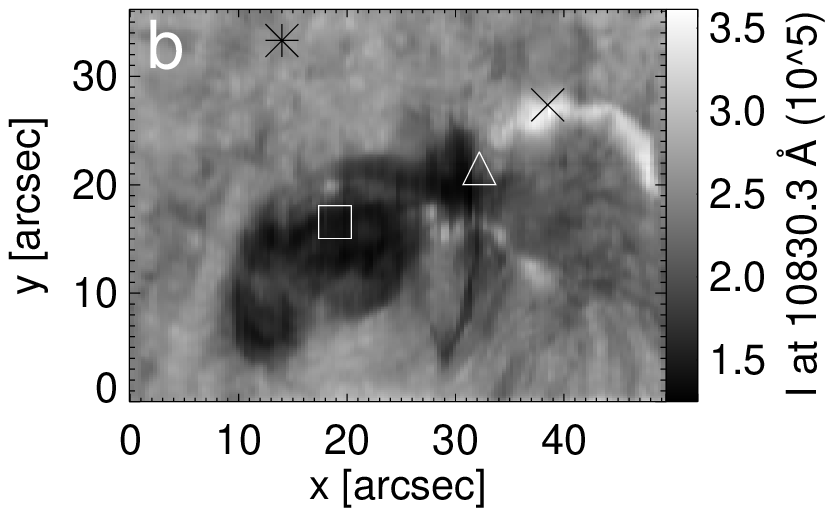}
\caption{Intensity maps at different wavelengths of the scanned
portion of active region NOAA 10763 (24$^\circ$ W, 14$^\circ$ S).
Left (a): Continuum. Right (b): \ion{He}{i} line core of the Tr2,3
component. The four symbols overplotted on the right image refer to
the line profiles shown in Fig.~\ref{fig:diff_prof}.}
\label{fig:maps}
\end{figure*}

On 2005 May 18 we observed the active region NOAA 10763, located at
$24^\circ$ W, $14^\circ$ S on the solar disk, which corresponds to a
cosine of the heliocentric angle, $\mu=0.9$. During the observations
a flare of GOES class C2.0 erupted in the western part of the active
region. A filament was present in the active region close to and
partly overlying the flare ribbons, which was activated by the
flare. \\
The analyzed data were recorded with the second generation Tenerife
Infrared Polarimeter TIP-II \citep{collados} mounted at the 70~cm
aperture Vacuum Tower Telescope (VTT) at the Teide observatory in
Tenerife. The TIP-II instrument allows for a spectral window of
11~{\AA} around the \ion{He} 10830~{\AA} line to be recorded with a
wavelength dispersion of 11~m{\AA} per pixel. The wide wavelength
range turned out to be absolutely necessary to cover the full range
of possible He absorption signatures observed during the evolution
of the filament. As we shall show below in Sect.~\ref{sec:results},
some of the \ion{He}{i} profiles we observed display absorption at
redshifts corresponding to as much as 100~km/s. This is considerably
higher than the highest redshifts measured by \citet{aznar}, whose
exhaustive analysis of supersonic downflows seen in the \ion{He}{i}
lines was limited by the spectral range available to the much
smaller TIP-I detector \citep{martinez}. \\
The spectral window of TIP-II (10825-10836~{\AA}) contains
photospheric lines of  \ion{Si}{i} at 10827.09~{\AA},  \ion{Ca}{i}
at 10829.27 and 10833.38~{\AA} and  \ion{Na}{i} at 10834.85~{\AA},
the chromospheric \ion{He}{i} multiplet and two telluric blends at
10832.11~{\AA} and 10833.90~{\AA} (see Table~\ref{tab:righe}). \\
The filament was scanned in steps of 0.35" perpendicular to the slit
orientation ($180.51^\circ$ with respect to the solar N-S
direction), from 14:38:29 to 15:02:26 UT, providing a
map of the region of $36.5\times25$~Mm$^2$. \\
Figure~\ref{fig:maps} shows Stokes $I$ maps of the observed region
obtained in the continuum at 10825.8~{\AA} (a) and in the core of
the Tr2,3 component of the  \ion{He}{i} triplet (b). The slit is
oriented along the $y$-axis and the direction of the scan is along
the $x$-axis. Therefore, along the $x$-axis both spatial and
temporal information are mixed. In addition to a few small
pores, Fig.~\ref{fig:maps}a clearly shows the granulation across the
whole image, suggesting that the seeing conditions during the scan
were particularly good and stable, allowing on average a resolution
of 1". The resolution was determined by performing a
Fast Fourier Transform of the average continuum image. Note that the
diffraction-limited performance of the VTT at this wavelength
corresponds to a resolution of 0.39". At some locations the
\ion{He}{i} profile is so broad that no pure continuum is present on
the detector. These locations appear as dark cloud-like features in
Fig.~\ref{fig:maps}a. In Fig.~\ref{fig:maps}b a
strong absorption from the filament material is clearly visible in
the He~\textsc{i} lines at $9"<x<33"$, scanned from $\sim$ 14:43 to
14:53 UT. \\
In the last part of the map ($33"<x<49"$), at $y\approx19"$-28",
during the scan of a flare ribbon, the absorption in the
 \ion{He}{i} lines is weaker than in quiet Sun regions and the
spectra also show in part emission profiles \citep[visible as
brightenings in the image; cf.][]{lagg1}. In this region the
\ion{H}{$\alpha$} slit-jaw images that were simultaneously recorded
at the VTT display flare ribbons as well.

\section{Analysis of the Stokes profiles}

\begin{figure}
\centering
\includegraphics[clip=true,width=8cm]{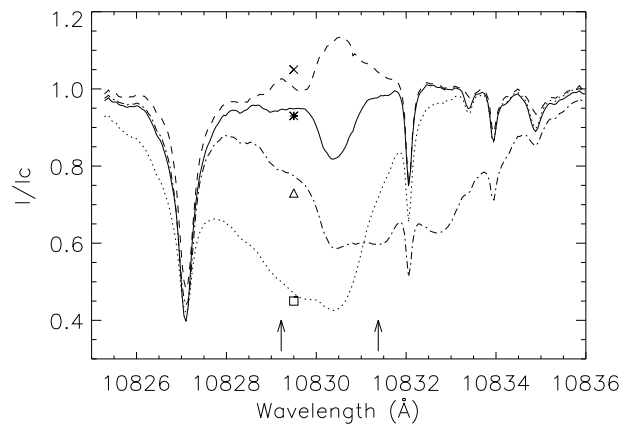}
\caption{Observed Stokes $I$ profiles at different locations,
indicated by the different symbols in Fig.~\ref{fig:maps}b. Solid
line (star symbol): quiet Sun profile. Dashed line (cross):
\ion{He}{i} emission profile. Dash-dotted line (triangle): very
broad  \ion{He}{i} profile indicating multiple redshifted
atmospheric components. Dotted line (square): deep absorption in the
He lines, that is in part strongly blueshifted. The arrows mark the
wavelengths at which images are shown in Fig.~\ref{fig:flows}.}
\label{fig:diff_prof}
\end{figure}
\begin{figure*}
\centering
\includegraphics[clip=true,width=8cm]{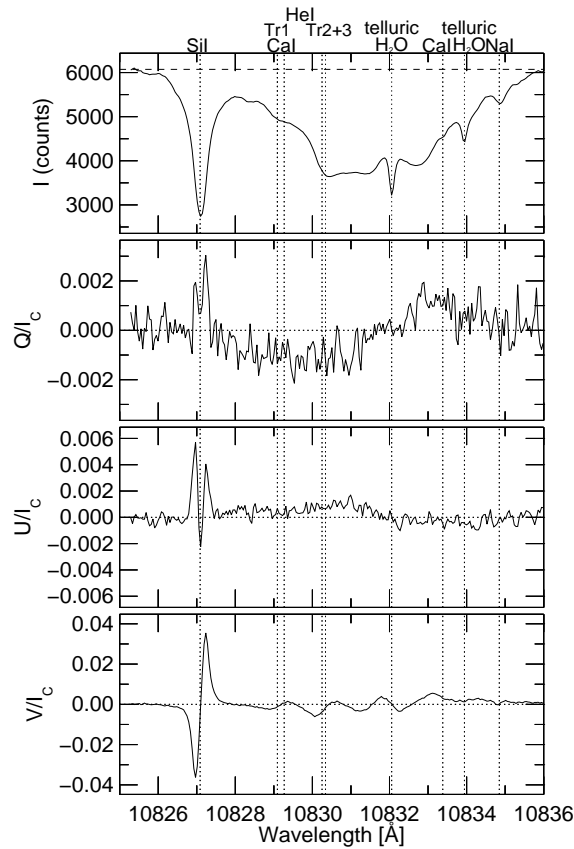}
\caption{Example of an observed averaged
Stokes vector (dash-dotted profile of Fig.~\ref{fig:diff_prof}). The
unusually broad Stokes $I$ profile and the corresponding signatures
of the $V$ profile suggest that the He triplet samples multiple atmospheric
components. The vertical dotted lines
mark the rest wavelengths of the different lines that are present
in the observed wavelength range and are identified at the top of the figure
and listed in Table~\ref{tab:righe}.}
\label{fig:fai_plot}
\end{figure*}

\begin{figure*}
\centering
\includegraphics[clip=true,width=6.86cm]{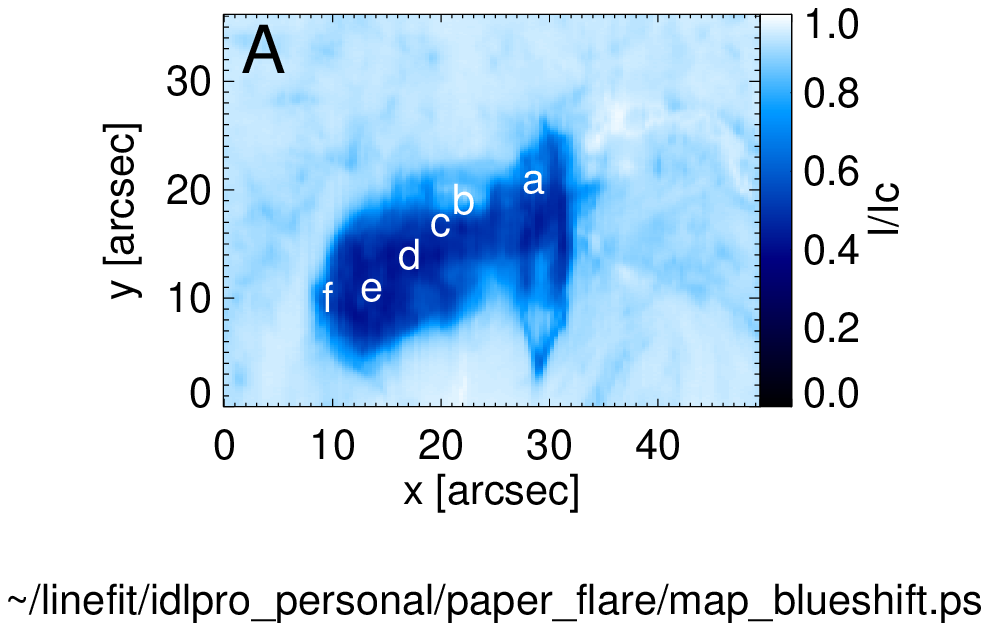}
\includegraphics[clip=true,width=6.86cm]{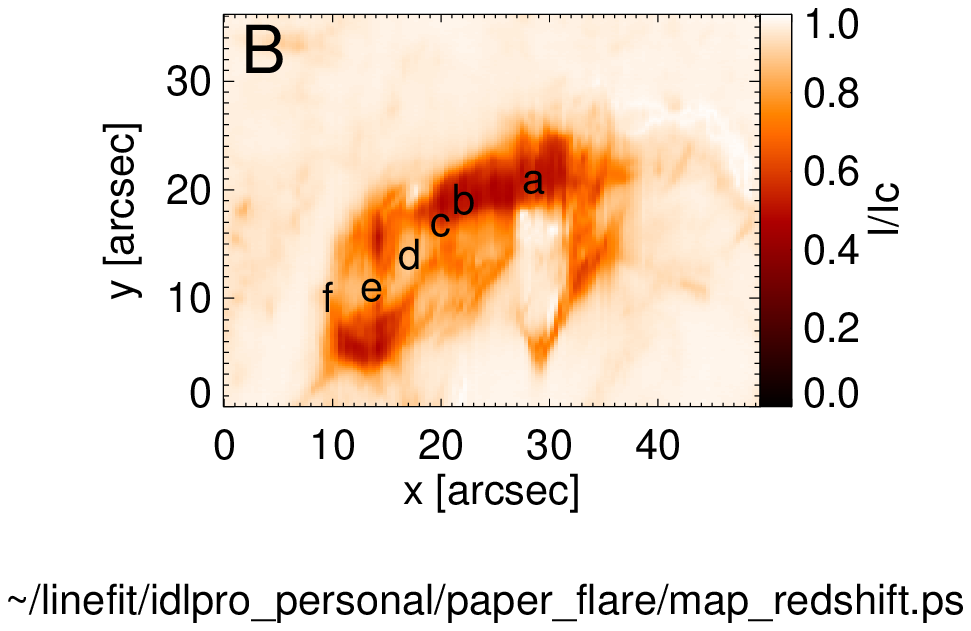}
\caption{Intensity maps of the scanned active region obtained at
different wavelength positions indicated by arrows in
Fig.~\ref{fig:diff_prof}. Left (A): At -30~km/s with respect to the
central rest position of the Tr2,3 component of the He triplet.
Right (B): At +30~km/s with respect to the central rest position of
the Tr2,3 component of the He triplet. The different letters refer
to the line profiles shown in Fig.~\ref{fig:profiles}.}
\label{fig:flows}
\end{figure*}

\begin{figure*}
\centering
\includegraphics[clip=true,width=6.86cm]{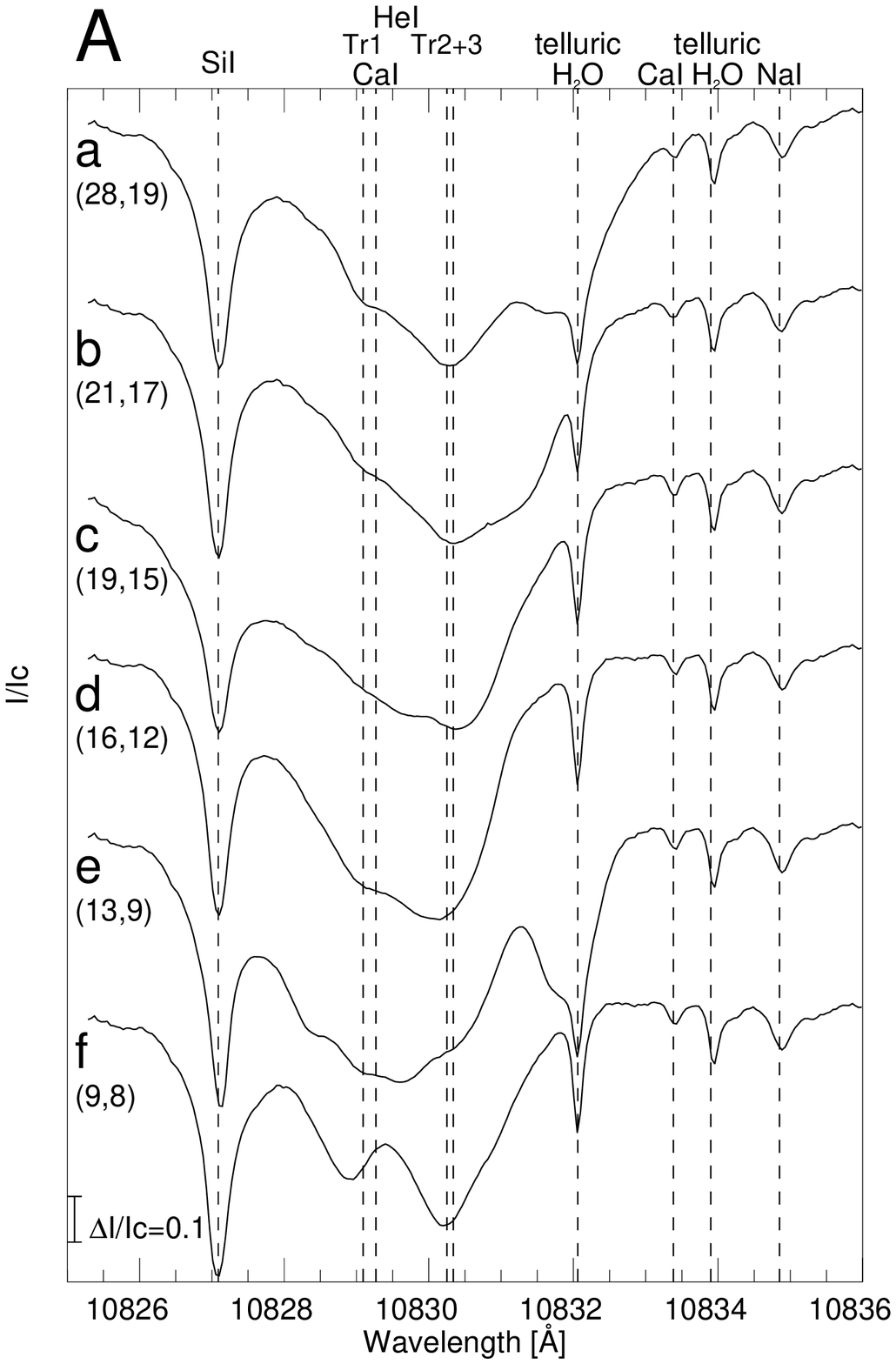}
\includegraphics[clip=true,width=6.86cm]{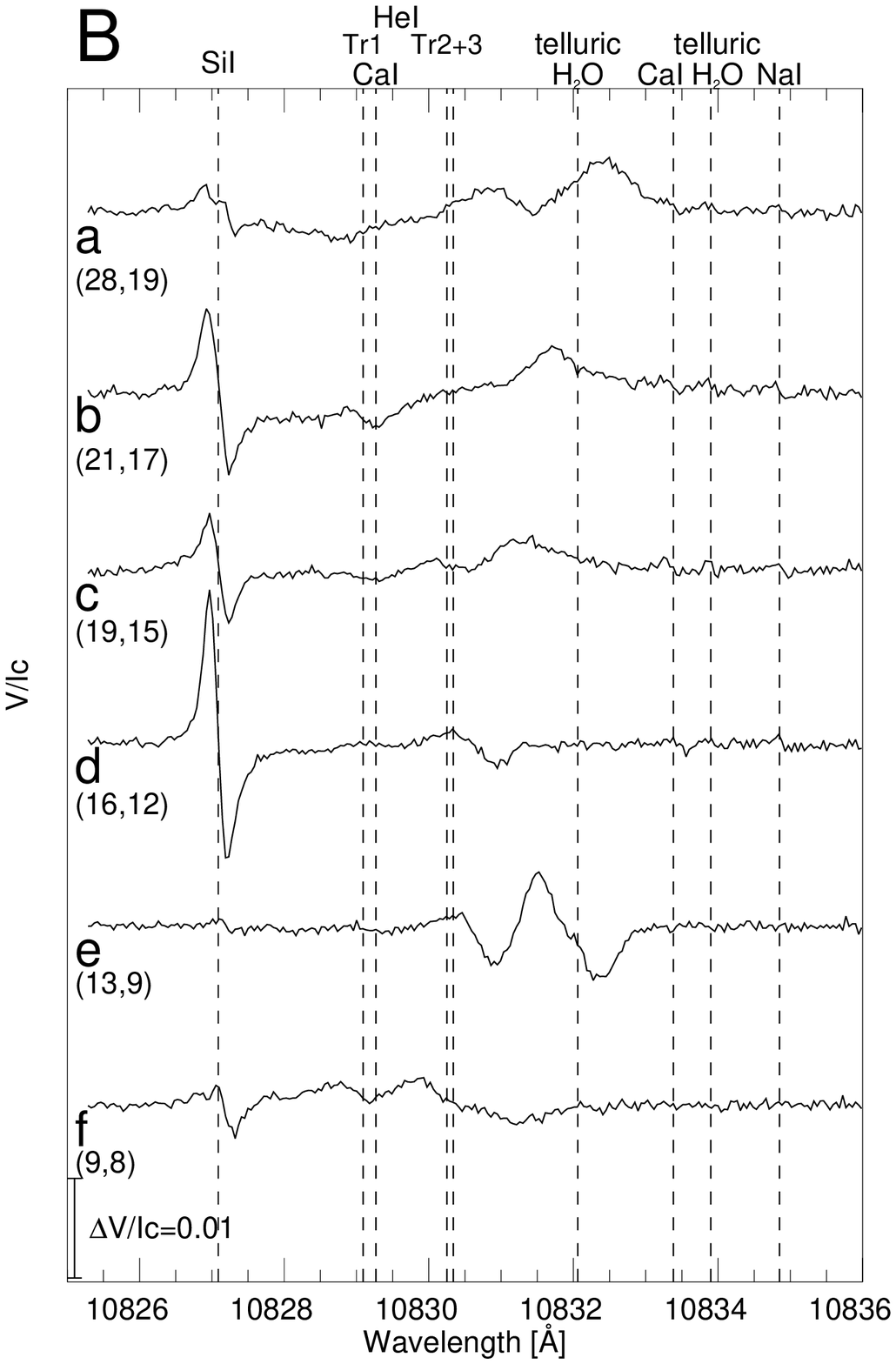}
\caption{Observed averaged Stokes $I$ (A) and
$V$ (B) profiles at the locations marked by letters in
Fig.~\ref{fig:flows}. The dashed, vertical lines mark the central
position of the spectral lines in the wavelength field-of-view
of the TIP-II instrument. The (x,y) coordinates in arcseconds
are written next to each profile.}
\label{fig:profiles}
\end{figure*}

In Fig.~\ref{fig:diff_prof} four observed Stokes $I$ profiles
normalized to the continuum intensity are plotted as an example of
the unusually wide range of profiles observed. These spectra and all
the spectra we show here are binned over four spectral
pixels (resulting in a wavelength dispersion of 44~m{\AA}) and eight
spatial pixels (resulting in a spatial pixel size of
$0.7\times0.7$~arcsec$^2$, a single pixel being
$0.35\times0.18$~arcsec$^2$) in order to increase the
signal-to-noise ratio to $\sim$ 4000. The solid line represents a
profile observed in a comparatively quiet region of the map (at the
location of the asterisk in Fig.~\ref{fig:maps}b), the dashed
spectrum shows a \ion{He}{i} profile in emission observed in a flare
ribbon (cross in Fig.~\ref{fig:maps}b). Note that the \ion{Si}{i}
line is weakened at this location. The dash-dotted line (triangle in
Fig.~\ref{fig:maps}b) shows a very broad absorption in the
\ion{He}{i} lines observed in the filament. Significant absorption
covers the wavelength range from 10829~{\AA} to 10835~{\AA}, with
weak absorption to the very edge of the detector
at 10836~{\AA}. The Stokes $I$ profile exhibits multiple minima that are not
corresponding to the rest wavelengths of lines normally found in the
Sun's spectrum, nor to telluric absorptions. The multiple absorption dips in
the profile suggests multiple atmospheric components with
a shift in the \ion{He}{i} absorption caused by different velocities.
For the dash-dotted profile
these components are nearly at rest or are redshifted (see also
Fig.~\ref{fig:fai_plot}). The maximum redshift corresponds to a
downflow velocity along the line-of-sight of $\sim100$~km/s. The
dotted line (square in Fig.~\ref{fig:maps}b) is an example of a
particularly deep absorption in the \ion{He}{i} line, which becomes
as deep as the strong \ion{Si}{i} line in the quiet Sun. In
addition to this strong unshifted absorption there is a broad
blueshifted absorption that blends with the \ion{Si}{i} 10827~{\AA}
line and extends even bluewards of it, revealing
rapidly upward propagating chromospheric gas. The maximum blueshift
corresponds to a velocity along the line-of-sight of $\sim60$~km/s.
As we already pointed out, the shown spectra are always binned, but
multiple components are also seen in the single
profiles forming the composite. \\
In Fig.~\ref{fig:fai_plot} we plot the (averaged) full Stokes vector for the
dash-dotted profile of Fig.~\ref{fig:diff_prof} (position $x=32"$,
$y=22"$ in Fig.~\ref{fig:maps}). The polarized Stokes
parameter displaying the highest signal-to-noise ratio is Stokes
$V$. It displays a significant signal over nearly the entire
wavelength range over which Stokes $I$ exhibits a strong
absorption. The complexity of the Stokes $V$ profile confirms that
the broad \ion{He}{i} absorption is most likely caused by
multiple atmospheric components harboring gas that  flows at
different speeds. It also indicates that at least a part of the
downflowing gas is located in magnetized regions of the solar
atmosphere. Stokes $Q$ and, to a lesser extent, Stokes $U$ also
display a signal over a broad wavelength range, although the signal
is weak and consequently the signal-to-noise ratio is fairly low. In
Fig.~\ref{fig:fai_plot} the wavelength positions at rest of the
different lines present in the observed wavelength range are
indicated by the vertical dotted lines (see also Table~\ref{tab:righe}). \\
In order to investigate how the multiple He components are related
with the motion of the filament material, we
show in Fig.~\ref{fig:flows} two intensity maps of the scanned active region 
at wavelength
positions $-1.08$~{\AA} (A) and $+1.08$~{\AA} (B) with respect to
the rest wavelength of the Tr2,3 component of the He lines. The
wavelengths at which the images are shown in Fig.~\ref{fig:flows}
correspond to up- and downflows of 30~km/s, respectively, and are
marked by arrows in Fig.~\ref{fig:diff_prof}. From these maps we can
see that strong blue- or redshifted line profile components are
present over practically the whole filament, suggesting strong dynamics
throughout this active filament. These dynamics
can be very complex, because we note a number of locations at which
profiles with strong blue- and redshifts coexist within a
resolution element. Consequently, upflows and downflows are present
at the same spatial position, although with this simple analysis alone
we cannot rule out that these profiles are simply strongly
turbulently broadened, or a strong Tr1 mimics a blueshift. Indeed,
a similar anomalous broadening was observed by \citet{lopez}
in the \ion{He}{i}{D$_3$} line in spicules. They found that a
distribution of velocities with FWHM=50~km/s was required to
reproduce the observed broadening. \\
In Fig.~\ref{fig:profiles} we present some examples of observed
averaged Stokes $I$ (A) and Stokes $V$ (B) profiles at different
positions marked by the letters a-f in Fig.~\ref{fig:flows}. They
stretch approximately from one end of the filament to the other and
display a gradual change from mainly redshifted (a) to mainly
blueshifted (f) profiles. All these profiles are characterized by
multiple atmospheric components of the  \ion{He}{i}
lines. In some of them, most prominently in c and e, blueshifted and
redshifted components coexist. Note also the changing ratios of the
\ion{Si}{i} and the \ion{He}{i} Stokes $V$ profile amplitudes. At
some locations, notably b and c, the photospheric and chromospheric
Stokes $V$ profiles have opposite polarities. This indicates that
the magnetic structure changes quite dramatically between the
photosphere and the chromosphere, which is not unusual for filaments
\citep{leroy,low}. We also have evidence of different He components
showing emission and absorption within the same profile (not shown
in this paper), which are generally co-located with the bright
\ion{H}{$\alpha$} flare kernels, but do not necessarily appear
bright in Fig.~\ref{fig:maps}b.


\subsection{Stokes inversion}\label{sec:inversion}

\begin{figure}
\centering
\includegraphics[clip=true,width=8cm]{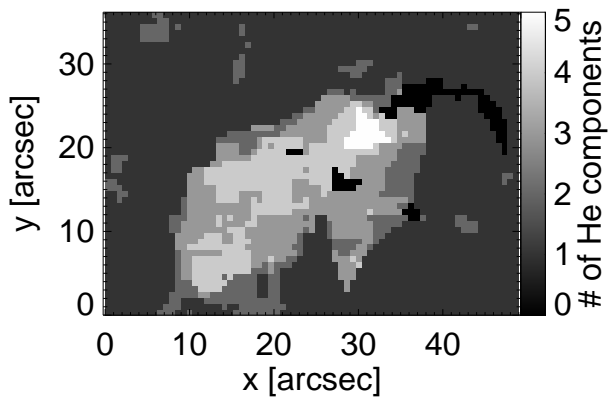}
\caption{Number of He components needed to fit the observed profiles
for each pixel of the map. In the map, the pixels containing
profiles that were not inverted are black (mainly those in which the
He profile is in emission).} \label{fig:numcomp}
\end{figure}

We analyze the spectropolarimetric observations with the numerical
code H{\footnotesize E}LI{\footnotesize X}\footnote{The inversion
code H{\footnotesize E}LI{\footnotesize X} is freely  available for
the solar community. Please contact A. Lagg to obtain a copy.} for
the synthesis and inversion of Stokes profiles in a Milne-Eddington
atmosphere \citep{lagg} extended to include the Paschen-Back effect
\citep{sasso,socasa}. H{\footnotesize E}LI{\footnotesize X} allows for
inversions of a single spectrum using multiple atmospheric
components, so that atmospheric parameters like the magnetic field
vector and the line-of-sight (LOS) velocity are retrieved for each
component. The free parameters tuned by the code to fit the observed
Stokes profiles are the magnetic field strength $B$ and direction
(inclination angle $\gamma$ and azimuthal angle $\chi$), the
line-of-sight velocity $v_{LOS}$, the Doppler width
$\Delta\lambda_D$, the damping constant $\Gamma$, the ratio of the
line center to the continuum opacity $\eta_0$ and the gradient of
the source function $S_1$. For more information on these parameters
we refer to \citet{lagg}. A range for every parameter to be fitted
is specified to guarantee that the result of the
inversion stays within the regime of physically useful solutions. \\
The inversion includes a scattering polarization correction, a
treatment of the Hanle effect for a special case. It is only valid
at disk center and for horizontal fields and it only improves the
azimuth retrieval \citep{Trujillo_nature}. Indeed, in the particular
case of a LOS-direction perpendicular to the solar surface (disk
center) and a magnetic field direction parallel to the solar surface
(horizontal field) the polarization signal owing to the Hanle effect
is oriented along the magnetic field direction: $\tan(2\chi)=(U/Q)$. \\
The details of the input atmospheres depend on the number of He
components necessary to reproduce the observed profiles. One
parameter, the filling factor $\alpha_i$, defines the contribution
of a given atmospheric component $i$ to the total observed He
profile. The sum of the filling factors of all atmospheric
components within a resolution element is required to be unity,
$\sum_i \alpha_i=1$. The maximum number of He atmospheric components
we need to reproduce the profiles in the scan is five (see
Fig.~\ref{fig:numcomp}). This is considerably more than in
previous studies involving the He line, and we need to demonstrate
that these fits are reliable. Additionally, we need to determine the
maximum amount of information that can be reliably gathered from
these complex inversions. \\
We have tested that we do not employ more components than necessary
by first fitting each profile with a smaller number of atmospheric
components and consequently free parameters. Only a significant
improvement of the fitness function $f$ \citep[defined in][]{lagg},
which gives the goodness of the fit, warrants the use of more
components or free parameters. Usually, an insufficient number of
free parameters results in a poor fit for parts of the observed line 
profile. \\
For some extreme profiles we face the problem of a huge number
of free parameters to be fitted. The spectral region covered by the
He absorption signature of strongly broadened or shifted profiles
also contains four photospheric lines of \ion{Si}{i}, \ion{Ca}{i}
(two lines) and \ion{Na}{i} and two telluric blends (see
Table~\ref{tab:righe}). We have to fit these lines as well to
obtain unbiased fits of the \ion{He}{i} absorption, which often blends
with some or all of these lines. Because the
photospheric magnetic and velocity vector in general
significantly differs from the chromospheric parameters, this increases the
total number of atmospheric components and requires additional
line--specific parameters of each line to be determined. For
simplicity, each spectral line was assigned its own atmospheric
component. The number of free parameters was limited by coupling
them appropriately (see Sect.~\ref{sec:coupling}). To obtain a good
fit of the strong photospheric \ion{Si}{i} line at 10827.09~{\AA},
we had to consider two atmospheric components: a magnetic component
and a field-free stray-light component. Only this combination could
satisfactorily reproduce both the line core and wings of the \ion{Si}{i}
line. The two telluric blends result in eight additional free
parameters, originating from fitting two Voigt profiles to the
telluric lines. The parameters needed to fit the telluric blends
could be kept in a very narrow range over the whole map, introducing
no decrease in stability of the convergence of the inversion. \\
There are profiles in the scan where the He lines show emission
features. These profiles were excluded from further analysis. Even
though the H{\footnotesize E}LI{\footnotesize X} code can fit
emission profiles by setting the gradient of the source function
$S_1$ to a negative number, the analysis technique is not optimized
to produce reliable results. Moreover, in some spatial pixels the He
emission profiles coexist with blue- or redshifted He absorptions,
making the retrieval even more complicated. In the map shown in
Fig.~\ref{fig:numcomp}, the pixels containing profiles that were not
inverted are black.

\subsection{Parameter coupling}\label{sec:coupling}

In order to decrease the number of free parameters and thus to
enhance the stability of the fitting procedure, we coupled some
atmospheric parameters between the atmospheric components, so that a
given atmospheric parameter was forced to have the same (but not
prescribed) value in all atmospheric components. Coupling between
parameters of different components was also used as a means to
impose that all lines from the same atmospheric layer (in particular
the photosphere) must be formed in the same atmosphere. We also
narrowed the ranges of possible solutions for some other parameters,
which also helped to ensure the uniqueness of the solution. \\
In particular, we coupled the magnetic field vector ($B$, $\gamma$
and $\chi$) and the gradient of the source function $S_1$ between
the four photospheric lines, assuming that their values stay
approximatively constant. This is consistent with the assumption
that the four lines are formed at similar heights and can be
represented by a single atmosphere. \\
Among the photospheric lines, the \ion{Si}{i} line is the strongest
one with the best signal-to-noise ratio and shows clear signatures in
the Stokes profiles. Changes in the atmospheric parameters will have
a larger influence on the Stokes profiles of the Si line than on the
weaker photospheric lines. Therefore, the photospheric parameters
are mainly determined by the shape and strength of the Si Stokes
profiles. Then, we made an inversion run for the whole map, fitting
only the photospheric and the telluric lines. From the results of
the fit, assuming that these lines are not disturbed by the filament
activity happening at higher atmospheric layers, we could select a
narrower input range for the Doppler width $\Delta\lambda_D$ of the
photospheric lines and the width and the damping of the Voigt
profile for the telluric blends. The narrower input ranges are then
used for the inversion done including the He lines. This avoids that
the code broadens and shifts these lines by large amounts to mimic
one of the He components. \\
The different He components certainly have different LOS velocities,
but it is not a priori clear if they sample different magnetic
vectors (and if they do, whether there is sufficient information in the
measured Stokes parameters to distinguish between different magnetic
vectors affecting the various \ion{He}{i} absorption components).
Therefore, we made several test runs to decide whether to couple the
magnetic field vector or not (see Sect.~\ref{sec:results}). \\
We also coupled the gradient of the source function $S_1$ between
the He atmospheric components, and we could also assign a relatively narrow 
input
range for $v_{LOS}$ of each He component  from the visual analysis of the
observed profiles. Indeed, when the velocity
separation between the He profiles is distinct enough ($\sim 10$~km/s)
we can clearly distinguish the positions of the different minima in
the $I$ profile.

\section{Results}\label{sec:results}

\begin{table}
\caption{Values of the fitness functions $f$ returned by the fits to
25 observed averaged profiles.}\label{tab:ff} \centering
\begin{tabular}{c|ccccc}
  \hline\hline
  Pixel x/y & 78-79 & 80-81 & 82-83 & 84-85 & 86-87 \\
  (arcsec)  & (27.3)& (28.0)& (28.7)& (29.4)& (30.1)\\
  \hline
  104-107 & $f_1=1.44$ & 1.42 & 1.53 & 1.54 & 1.29 \\
  (17.7)& $f_2=1.48$ & 1.71 & 1.75 & 1.58 & 1.17 \\
        & $f_3=1.15$ & 1.46 & 1.29 & 1.39 & 1.16 \\
  \hline
  108-111 & $f_1=1.95$ & 1.81 & 1.74 & 1.86 & 1.18 \\
  (18.4)& $f_2=1.91$ & 1.55 & 1.76 & 1.77 & 1.04 \\
        & $f_3=1.67$ & 1.48 & 1.48 & 1.69 & 1.08 \\
  \hline
  112-115 & $f_1=1.89$ & 1.68 & 1.73 & 1.25 & 1.50 \\
  (19.0)& $f_2=2.00$ & 1.63 & 1.60 & 1.28 & 0.94 \\
        & $f_3=1.78$ & 1.53 & 1.34 & 1.23 & 1.35 \\
  \hline
  116-119 & $f_1=1.59$ & 1.76 & 1.41 & 1.58 & 1.56 \\
  (19.7)& $f_2=1.47$ & 1.97 & 1.69 & 1.73 & 1.52 \\
        & $f_3=1.32$ & 1.80 & 1.36 & 1.46 & 1.40 \\
  \hline
  120-123 & $f_1=1.68$ & 1.98 & 1.51 & 1.54 & 1.76 \\
  (20.4)& $f_2=1.66$ & 1.88 & 1.51 & 1.38 & 1.60 \\
        & $f_3=1.61$ & 1.92 & 1.53 & 1.43 & 1.58 \\
 \hline
\end{tabular}
\end{table}

\begin{figure*}
\centering
\includegraphics[clip=true,width=6.86cm]{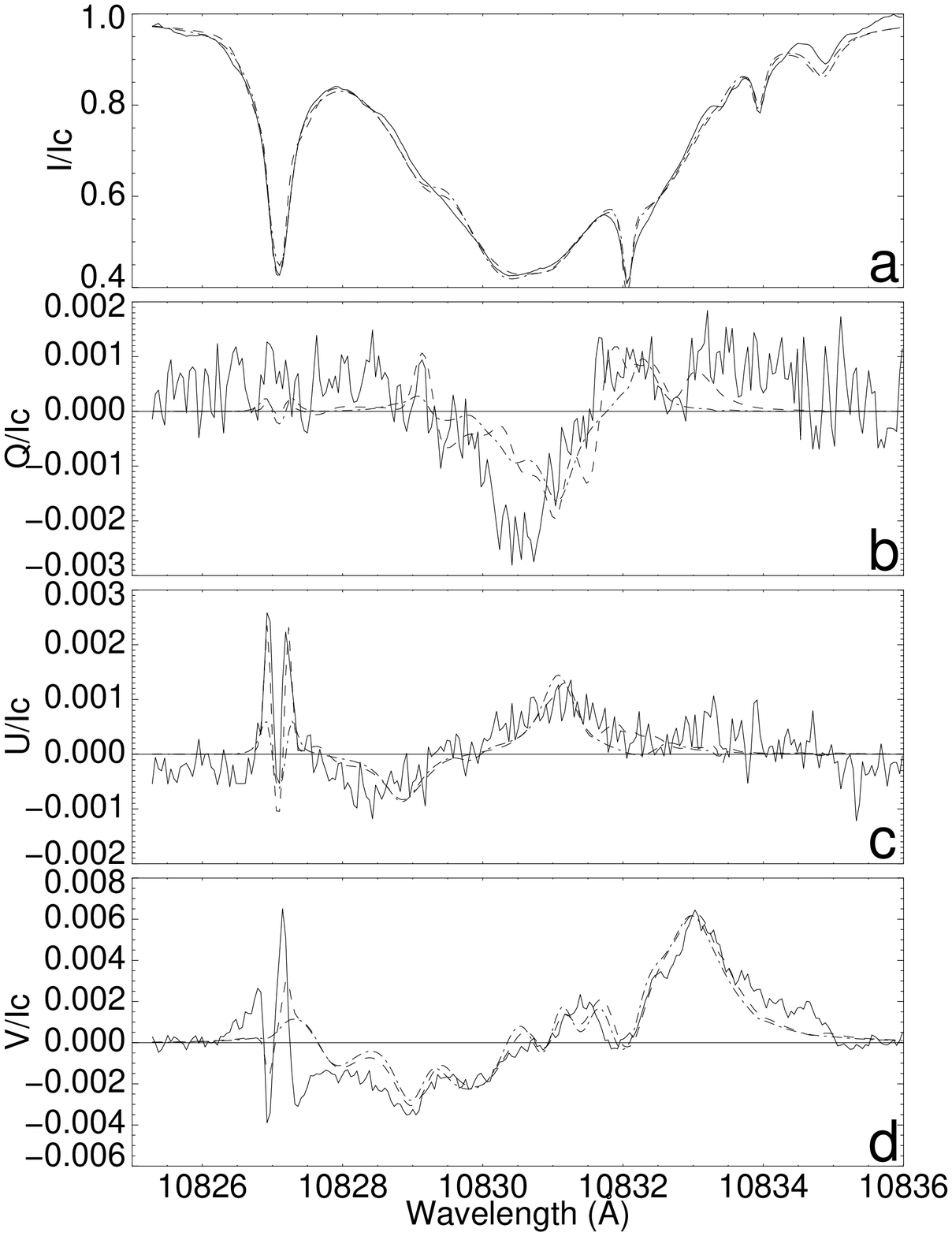}
\includegraphics[clip=true,width=6.86cm]{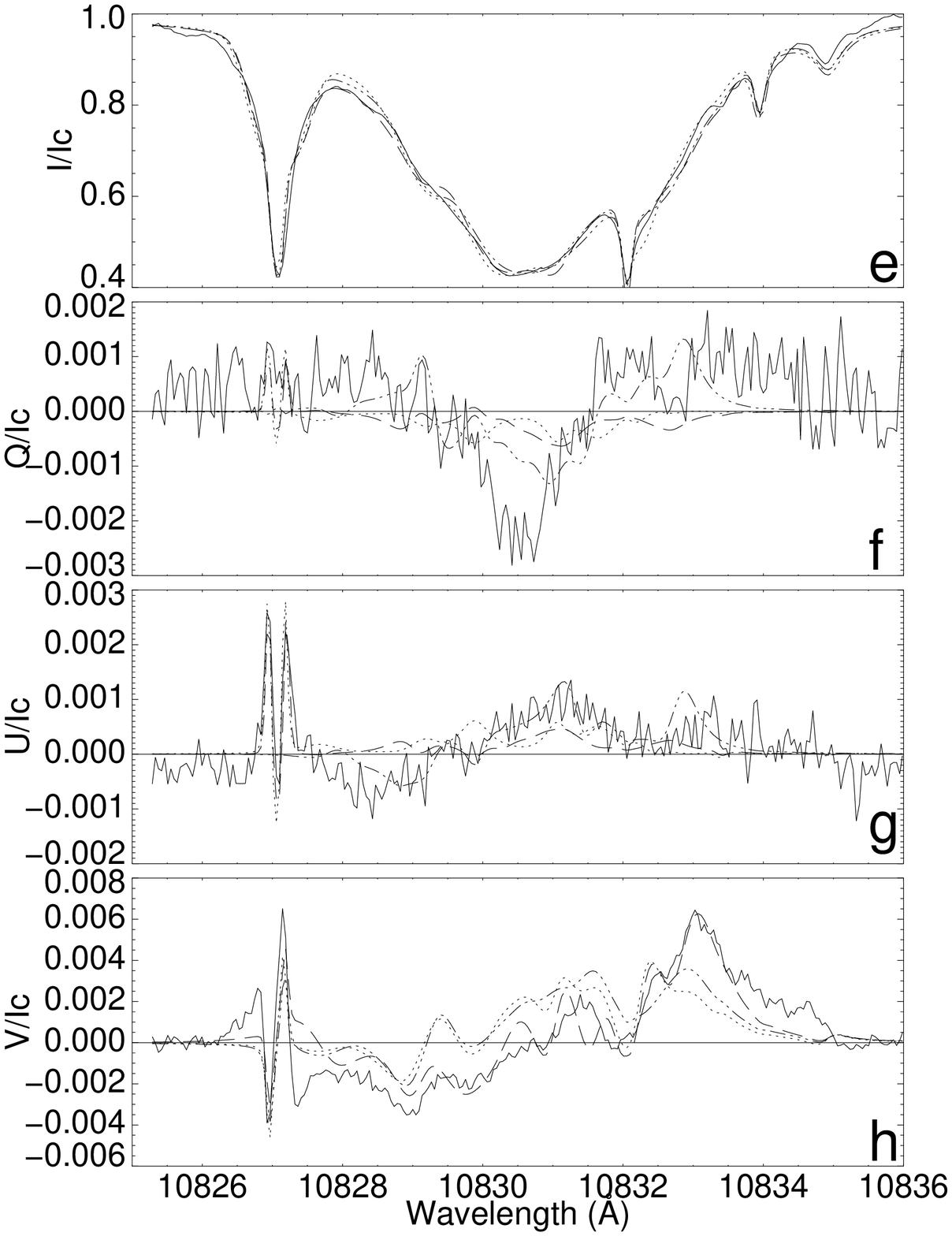}
\caption{Inversions of the Stokes $I$, $Q$, $U$, and $V$ averaged
profiles at the location (29", 19") of the map in
Fig.~\ref{fig:maps}. The fits (dashed, dash-dotted, dotted,
dash-dot-dotted and long dashed lines) to the observed profiles
(solid lines) are obtained by allowing five different atmospheric
components to describe the He I triplet. Left (a-d): the dashed
profile is obtained by leaving the magnetic field vector free
between the five He components while the dash-dotted profile is
obtained by coupling only the magnetic field strength. Right (e-h):
the dotted profile is obtained by coupling the magnetic field vector
between the five He components, coupling the magnetic field strength
and the inclination gives the dash-dot-dotted profiles, while
coupling the field strength and the azimuth leads to the long dashed
profiles.} \label{fig:profilo8288}
\end{figure*}

\begin{figure*}
\centering
\includegraphics[clip=true,height=10cm]{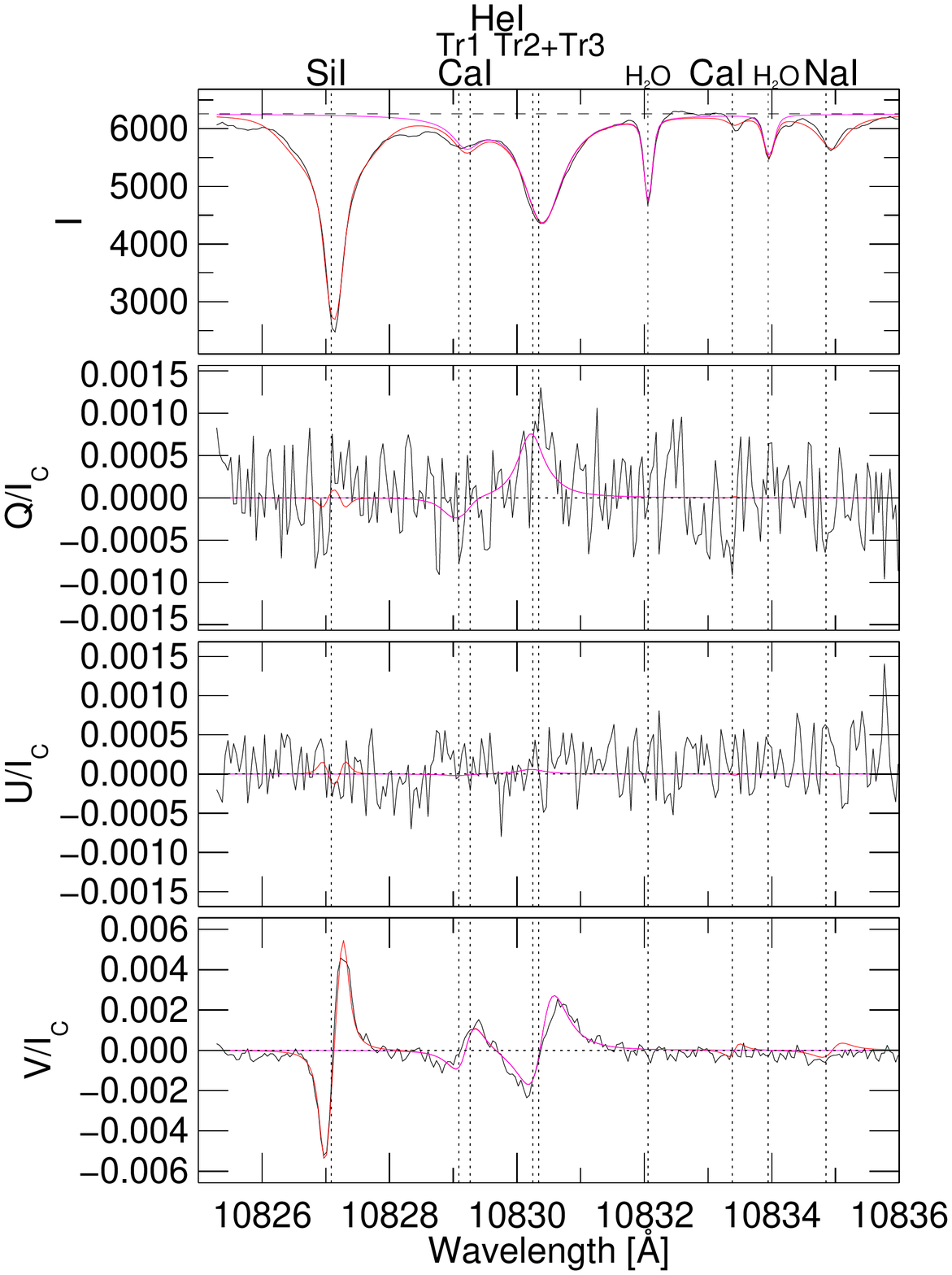}
\caption{Observed spatially averaged and
inverted Stokes $I$, $Q$, $U$ and $V$ profiles at the location (5",
13") of the map in Fig.~\ref{fig:maps}. The solid black lines are
the observed profiles, while the red lines are the fits to all
observed lines obtained imposing one atmospheric component to
describe the He I lines (magenta lines), and two for the
photospheric lines.}
\label{fig:profilo14132}
\end{figure*}

\begin{figure*}
\centering
\includegraphics[clip=true,height=10cm]{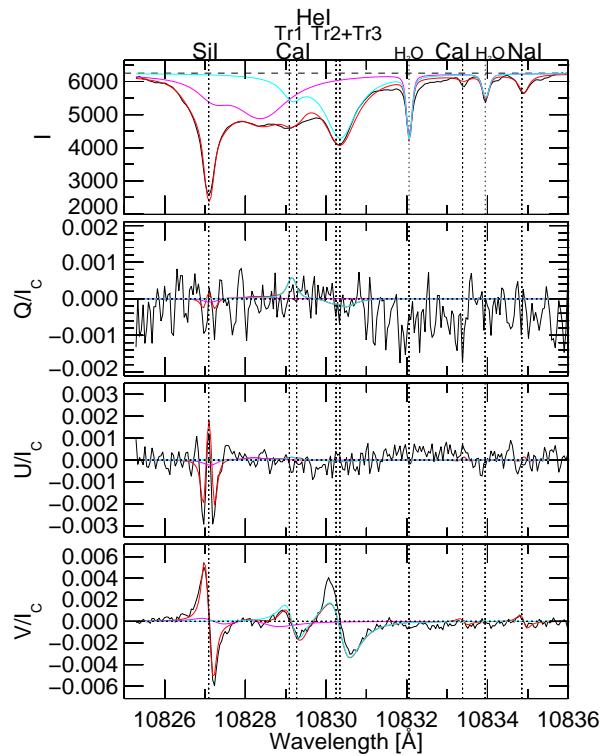}
\caption{Observed and inverted Stokes
$I$, $Q$, $U$, and $V$ profiles at the location (27", 9") of the map
in Fig.~\ref{fig:maps}. The solid black lines are the observed
profiles, while the red lines are the fits obtained imposing two
different atmospheric components to describe the He I lines (the
profiles of the individual components are given by magenta and
light blue lines).}
\label{fig:profilo78152}
\end{figure*}

\begin{figure*}
\centering
\includegraphics[clip=true,height=10cm]{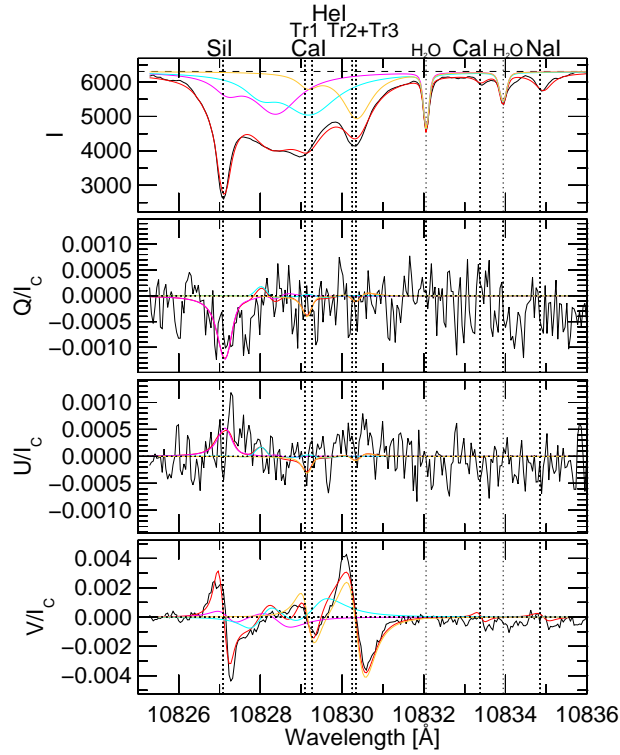}
\caption{Observed and inverted Stokes
$I$, $Q$, $U$, and $V$ profiles at the location (29", 13") of the map
in Fig.~\ref{fig:maps}. The solid black lines are the observed
profiles, while the red lines are the fits obtained by allowing three
different atmospheric components to describe the He I lines
(represented by magenta, light blue, and yellow lines).}
\label{fig:profilo82128}
\end{figure*}

\begin{figure*}
\centering
\includegraphics[clip=true,height=10cm]{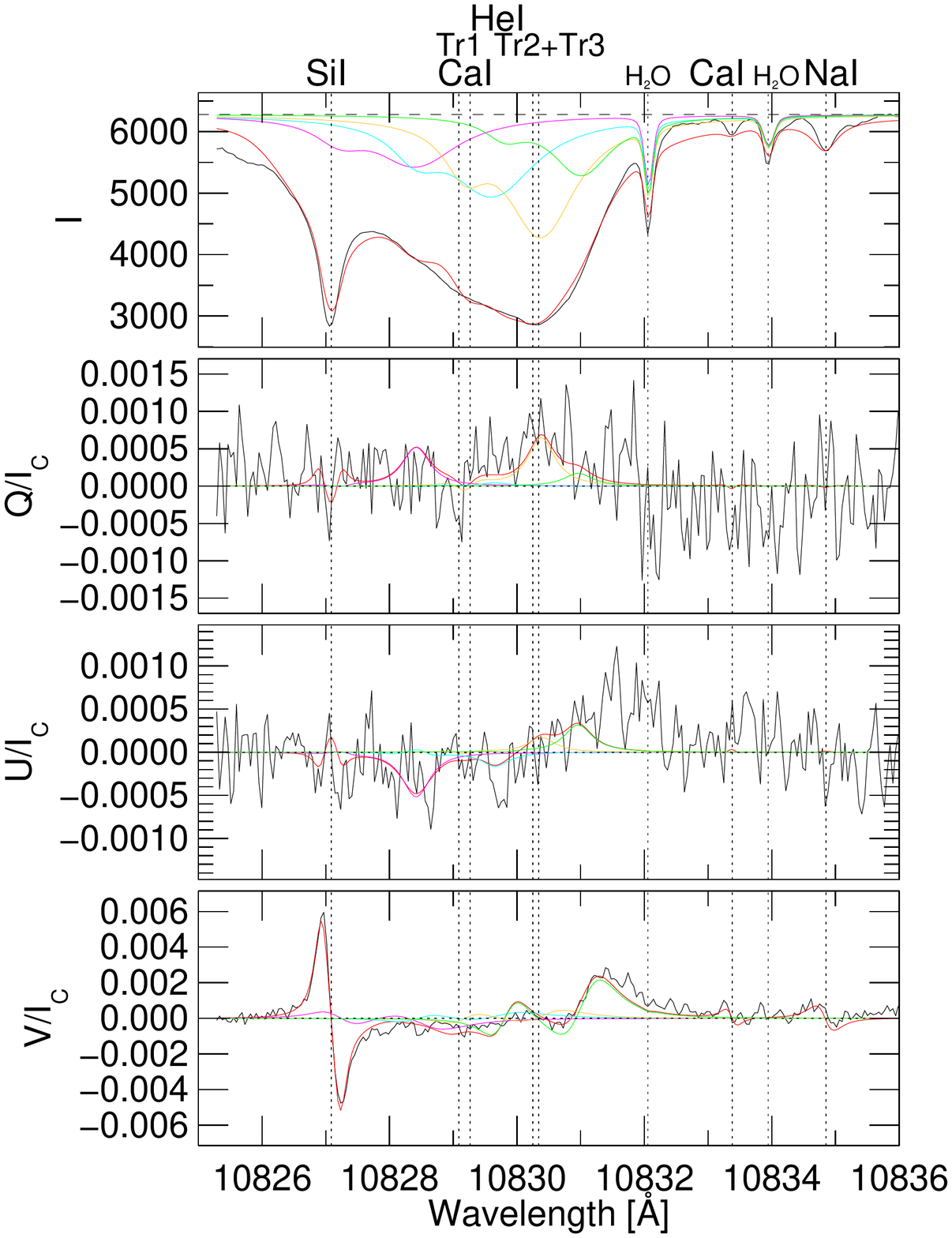}
\caption{Observed and inverted Stokes
$I$, $Q$, $U$, and $V$ profiles at the location (17", 16") of the map
in Fig.~\ref{fig:maps}. The solid black lines are the observed
profiles, while the red lines are the fits involving four different
atmospheric components to describe the He I lines (magenta, light
blue, yellow, and green lines).}
\label{fig:profilo50112}
\end{figure*}

\begin{figure*}
\centering
\includegraphics[clip=true,height=10cm]{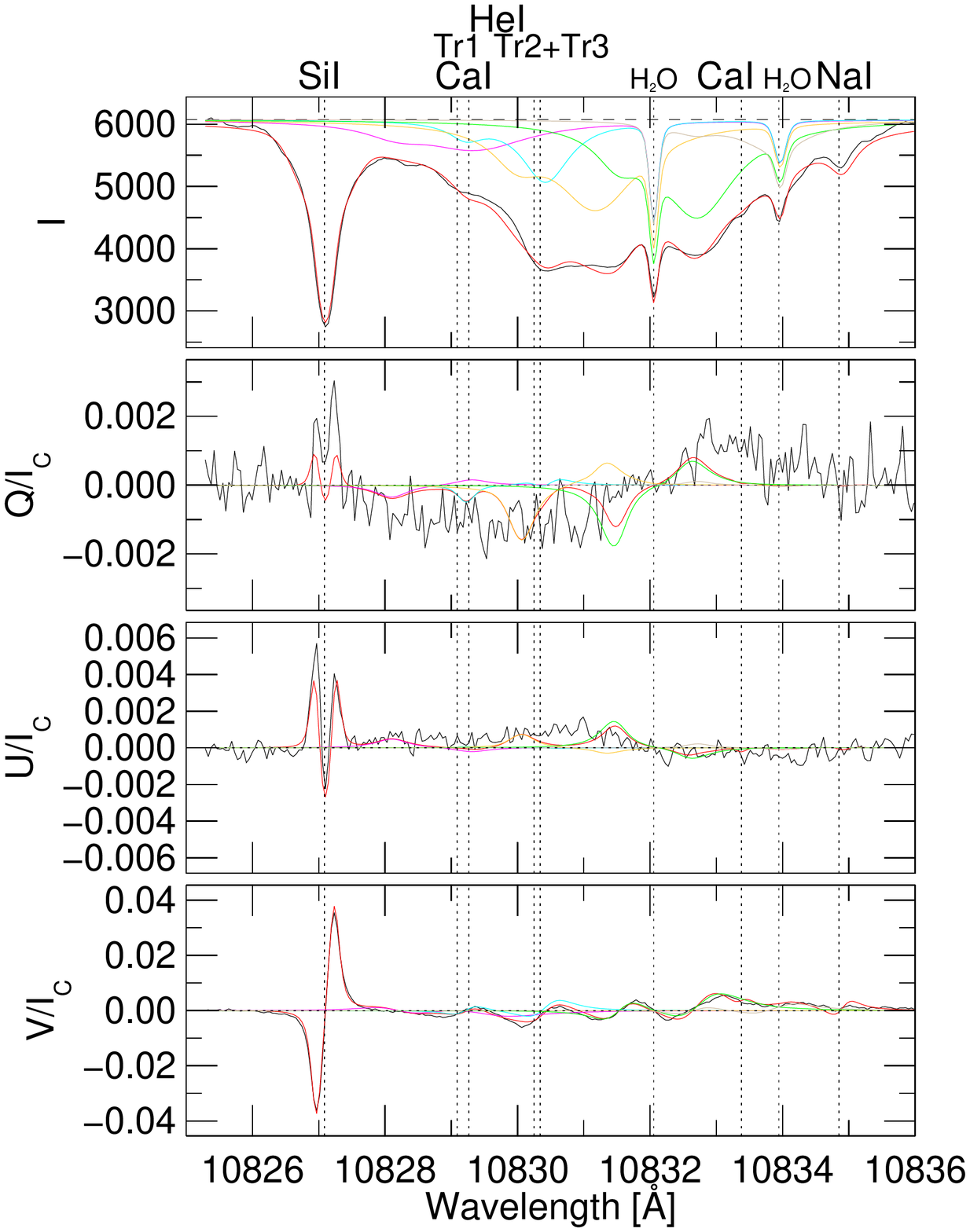}
\caption{Observed and inverted Stokes $I$,
$Q$, $U$, and $V$ profiles at the location (32", 21") of the map in
Fig.~\ref{fig:maps}. The solid black lines are the observed profiles,
while the red lines are the fits involving five different
atmospheric components to describe the He I lines (magenta, light
blue, yellow, green, and dark gray lines).}
\label{fig:profilo9284}
\end{figure*}

Test runs showed that the best fits to the observed Stokes profiles
are obtained by coupling only the magnetic field strength $B$
between the He components, leaving the two angles, $\gamma$ and
$\chi$, free. In Fig.~\ref{fig:profilo8288} we show as an example
the results of inversions of the observed averaged Stokes profiles
(solid lines) at the location (29", 19") of the map in
Fig.~\ref{fig:maps}. In Figs.~\ref{fig:profilo8288}(a-d) two fits to
the observed profiles (dashed and dash-dotted lines) are displayed.
Both are obtained by allowing five different atmospheric components
to describe the \ion{He}{i} lines. The dashed lines represent a fit
obtained by allowing the code to find a different magnetic field
vector for each atmospheric component used to describe the He
profiles. For the fit represented by the dash-dotted lines though,
we coupled the magnetic field strength between the components, while
leaving the angles describing the magnetic field direction
uncoupled. The average values of $f$ obtained from 100 inversion
runs of this particular observed profile are 1.62 for the dashed
line fit and 1.53 for the dash-dotted line fit. Both fits are good
and the fitness values are comparable, but for the dash-dotted
profile, the fit is obtained with 4 free parameters less. The
slightly higher value of the fitness of the dashed profile is not
enough to justify the use of so many more free parameters. In
Figs.~\ref{fig:profilo8288}(e-h) the same observed profile (solid
line) is plotted as in (a-d). The dotted line is obtained by
coupling the whole magnetic field vector between the five
components. The observed Stokes $Q$, $U$ and $V$ profiles cannot be
reproduced by this inversion ($f=0.89$). By coupling $B$ and $\chi$
(long dashed lines) we obtain a good fit only for $I$ and $V$
($f=1.26$) and by coupling $B$ and $\gamma$ (dash-dot-dotted lines)
we can fit $I$, $Q$, and $U$ reasonably well, but not $V$
($f=1.02$). In no case can we reproduce the
Stokes vector as a whole. \\
Stokes $Q$ of this observed profile shows a predominantly
single-lobe signature that suggests resonance scattering
polarization. The simple scattering polarization correction we used
and described in Sect.~\ref{sec:inversion} is not sufficient to fit
the strong $Q$ peak. An
improved treatment of the scattering polarization may help. \\
We carried out the same analysis for another 25 Stokes profiles
with 3, 4, and 5 He components in a region of $3.5\times1.8$~Mm$^2$
around the profile displayed in Fig.~\ref{fig:profilo8288}. The
values of the fitness function $f$ obtained from three different
types of fits (leaving the magnetic field free, $f_1$, coupling $B$,
$f_2$, and coupling $B$ and $\chi$, $f_3$) to the observed profiles
are grouped in Table~\ref{tab:ff}. The profiles are located at the
spatial pixels listed at the top (x) and on the left (y) of
Table~\ref{tab:ff}. These observed profiles can be reproduced almost
equally well either by coupling the magnetic field strength between
the He components or by leaving the full magnetic field vector free.
Significantly poorer fits to the full Stokes vectors are obtained by
coupling $B$ and $\chi$. Indeed, $f_3<f_1$ in 88\% of the cases and
$f_3<f_2$ in 80\% of the cases. $f_1$ is higher than $f_2$ in 56\%
of the cases, while in over 40\% of the cases $f_2>f_1$. The
difference between the two is insignificant and the marginally
higher value of $f_1$ is not enough to justify using a higher
number of free parameters. \\
Another test is based on the analysis of the errors in the retrieval
of the atmospheric parameters. The convergence of the genetic
algorithm used for the minimization of the merit function, Pikaia
\cite[]{pikaia}, does not depend on the choice of initial guesses
for the atmospheric parameters. The convergence to the global
minimum occurs on a random path after an infinite number of
iterations. We set an upper limit to the number of iterations (700)
in a way that the necessary computing time and the stability of the
resulting fitness, $f$, are in good balance. The variations of the
parameter values resulting from numerous inversion runs with an
upper limit to the number of iterations can be used as a direct
measure of the error in the retrieval of this parameter. This error
contains the intrinsic uncertainty in the parameter retrieval,
caused by the similarity of Stokes profiles for different parameter
values, and the error introduced by the inversion process itself. \\
The mean retrieved values of the atmospheric parameters with their
errors, as deduced from 100 inversion runs, are listed in
Table~\ref{tab:parameters8288} for each of the profiles shown in
Fig.~\ref{fig:profilo8288}. The five values per parameter correspond
to the different He atmospheric components. As pointed out earlier
both the dashed (no coupling) and the dash-dotted profiles (only $B$
coupled) provide almost equally good fits, but the latter is
obtained by an inversion with a lower number of free parameters. The
errors when $B$ is coupled are smaller than without coupling, in
particular the errors on $B$ and $\gamma$. This means that with a
lower number of free parameters the code is more stable. The
$v_{LOS}$ is always well retrieved with a significantly lower error
than the specified range for the $v_{LOS}$ of any given component
($\sim 10$~km/s). Both the magnetic field angles, $\gamma$ and
$\chi$, can be determined for the individual components. The error
in the inclination angle, $\gamma$, is sufficiently large that not
all components can be clearly distinguished from each other (e. g.
components 2 and 3 in Table~\ref{tab:parameters8288}), but the values
obtained are in general good enough to deduce the polarity of the
magnetic field. Regarding the azimuth, the signal-to-noise ratio in
the $Q$ and $U$ profiles is barely high enough to reliably retrieve
information on it. \\
The final three columns of Table~\ref{tab:parameters8288} list the retrieved
values with uncertainties for an increasing number of parameters
that are forced to have the same value in all atmospheric
components. Irrespective of whether we keep $B$ and $\gamma$
(dash-dot-dotted), $B$ and $\chi$ (long dashes) or $B$, $\gamma$ and
$\chi$ (dotted) coupled, the fitness $f$ is significantly lower than
if all parameters are uncoupled, or only $B$ is coupled.
Furthermore, the spread (uncertainty) in the retrieved parameters
tends to be larger in the last three columns, so that coupling
additional parameters does not lead to an increase in the
reliability of the retrieved solution, but exactly the opposite.
Furthermore, the retrieved values of the coupled parameters are
close to the $\alpha$-weighted averages of the uncoupled values of
the same parameter. The exception is $B$, for which the coupled
value is always lower. Based on the above test (and similar tests
applied to the Stokes profiles in other pixels) we inverted all the
observed profiles in the observed map by coupling the
magnetic field strength between the He components, and leaving
$\chi$ and $\gamma$ free (fitness $f_2$ in Table~\ref{tab:ff}). \\
Examples of the inversions of increasingly complex \ion{He}{i}
10830~{\AA} Stokes $I$, $Q$, $U$, and $V$ profiles are shown in
Figs.~\ref{fig:profilo14132}--\ref{fig:profilo9284}. In each plot,
the black lines are the averaged observed profiles, while the red
lines are the best fits to the observations obtained considering
one, two, three, four, and five atmospheric components of the He I
lines. These are the minimum numbers of atmospheric
components that are needed to obtain a satisfactory fit, both as
judged by eye and from a decrease of $f$ found by adding a further
component. Also plotted are the different He components (magenta,
light blue, yellow, green, and dark gray lines), whose sum gives the
red lines. The plotted contribution of each component is multiplied
by its filling factor. In Table~\ref{tab:parameters} we give the
mean retrieved values of the atmospheric parameters $B$, $\gamma$,
$\chi$, $v_{LOS}$ and $\alpha$, together with their errors. The
values listed in Table~\ref{tab:parameters} are averages of the
parameter values returned from 50 inversion runs, and the
uncertainties reflect the spread (standard deviation) of the
returned values. When multiple-component fits to the He line are
required, such as those displayed in
Figs.~\ref{fig:profilo78152}--\ref{fig:profilo9284},
the parameters retrieved for each component are given. Note that the
photospheric lines are simultaneously fitted by two atmospheric
components (one magnetic, one field free), as described in
Sect.~\ref{sec:inversion}. For each fit we also
give the value of the fitness function in the Table, $f \pm \Delta f$. \\
Figure~\ref{fig:profilo14132} displays a one-He-component fit to the
observed Stokes profiles. Similar quiet profiles are found in the
region of the scan outside the portion with the filament or
other activity (flare). In Table~\ref{tab:parameters} (first table)
the mean values and the errors of the retrieved atmospheric
parameters after 50 inversion runs are reported. Even with
this simple profile we have difficulties in retrieving the values of
the magnetic field angles, $\gamma$ and $\chi$, in particular $\chi$
with high accuracy. This is mainly because of the low signal-to-noise
ratio of the observed $Q$ and $U$ profiles and to a smaller extent
because of the already high number of free parameters to be fitted (the
inversion includes the He, the photospheric and the telluric lines).
Figures~\ref{fig:profilo78152} and \ref{fig:profilo82128} show a
two- and a three-component fit to other, more complex observed
profiles. The two profiles, observed close in position
and time, are somewhat similar, with the difference of a deeper He
blueshifted absorption in the latter profile. This additional
absorption requires an additional He component to be fitted. The
presence of another He component in the profile of
Fig.~\ref{fig:profilo82128} is also evident from the different shape
of the negative lobe in the \ion{Si}{i} $V$ profile. In both
profiles we find strongly supersonic blueshifted He components
(speeds in excess of -50~km/s) coexisting with a He component nearby
at rest. The fit to the observed profile shown in
Fig.~\ref{fig:profilo50112} requires an additional fourth
redshifted He component. The final fit shown in
Fig.~\ref{fig:profilo9284} requires five atmospheric components. Only
few other profiles (see Fig.~\ref{fig:numcomp}) need five components
to be reproduced, and these are all located around one end point of
the observed filament. This is also the location where we measure
the highest downflow velocities ($\sim 100$~km/s). It is
particularly gratifying to see that both the fitness $f$ and
the uncertainties of the retrieved parameters appear to be nearly
independent of the number of atmospheric components. This is
probably because profiles with more components tend to
have a larger equivalent width and individual components are well
separated in wavelength.

\begin{table*}
\caption{Mean values of the atmospheric parameters with the errors
returned by the fits displayed in
Fig.~\ref{fig:profilo8288}.}\label{tab:parameters8288} \centering
\begin{tabular}{c|ccccc}
  \hline\hline
  Parameter/Fit-profile  & Dashed     & Dash-dotted & Dash-dot-dotted & Long dashes & Dotted \\
  & No coupling& $B$ coupled & $B$,$\gamma$ coupled& $B$,$\chi$ coupled&$B$,$\gamma$,$\chi$ coupled\\
  \hline
  $B \pm \Delta B$ (G) &  $478\pm330$ & $228\pm53$ & $250\pm134$ & $234\pm55$ & $222\pm123$ \\
                       &  $264\pm215$ &          &        &              &   \\
                       &  $403\pm285$ &          &        &              &   \\
                       &  $307\pm159$ &          &        &              &   \\
                       &  $632\pm376$ &          &        &              &   \\
  \hline
  $\gamma \pm \Delta \gamma$ $(^\circ)$ & $47\pm25$ & $33\pm22$ & $122\pm23$ & $31\pm26$ & $124\pm22$ \\
                       &  $122\pm22$ &  $110\pm4$  &    & $110\pm5$  &      \\
                       &  $108\pm24$ &  $102\pm7$  &    & $102\pm7$  &      \\
                       &  $131\pm21$ &  $141\pm17$ &    & $143\pm17$ &      \\
                       &  $141\pm22$ &  $170\pm10$ &    & $169\pm12$ &      \\
  \hline
  $\chi \pm \Delta \chi$ $(^\circ)$ & $-29\pm34$  & $-18\pm48$ & $-30\pm43$ & $30\pm42$ & $27\pm44$\\
                       &  $46\pm70$ &  $52\pm62$  &  $60\pm53$   & &       \\
                       &  $58\pm28$ &  $48\pm32$  &  $50\pm32$   & &       \\
                       &  $7\pm16$  &  $0\pm13$   &  $1\pm13$    & &       \\
                       &  $8\pm33$  &  $-2\pm37$  &  $-3\pm37$   & &       \\
  \hline
  $v_{LOS} \pm \Delta v_{LOS}$ (km/s) & $-38\pm3$ & $-39\pm2$  & $-31\pm2$ & $-38\pm3$ & $-31\pm2$ \\
                       &  $1\pm1$  &  $1.1\pm0.7$ &  $0.9\pm1.0$  & $1\pm1$  & $0.9\pm1.0$\\
                       &  $21\pm1$ &  $21\pm1$    &  $21\pm1$     & $21\pm1$ & $21\pm1$\\
                       &  $53\pm1$ &  $53\pm1$    &  $53\pm1$     & $53\pm1$ & $53\pm1$\\
                       &  $68\pm2$ &  $69\pm2$    &  $68\pm2$     & $69\pm2$ & $68\pm2$ \\
  \hline
  $\alpha \pm \Delta \alpha$ & $0.10\pm0.03$ & $0.10\pm0.02$ & $0.09\pm0.02$ & $0.10\pm0.03$ & $0.10\pm0.03$\\
                       &  $0.33\pm0.06$ &  $0.32\pm0.04$ &  $0.29\pm0.04$ & $0.31\pm0.04$ & $0.27\pm0.05$\\
                       &  $0.20\pm0.05$ &  $0.19\pm0.03$ &  $0.19\pm0.04$ & $0.21\pm0.03$ & $0.21\pm0.03$\\
                       &  $0.25\pm0.06$ &  $0.24\pm0.03$ &  $0.27\pm0.04$ & $0.23\pm0.04$ & $0.26\pm0.04$\\
                       &  $0.12\pm0.05$ &  $0.14\pm0.03$ &  $0.16\pm0.04$ & $0.15\pm0.03$ & $0.16\pm0.03$\\
  \hline
  $f \pm \Delta f$ & $1.62\pm0.19$ & $1.53\pm0.16$ & $1.02\pm0.06$ & $1.26\pm0.09$ & $0.89\pm0.04$\\
  \hline
\end{tabular}
\end{table*}

\begin{table*}
\caption{Mean values of the atmospheric parameters with their errors
returned by the fits in
Figs.~\ref{fig:profilo14132}--~\ref{fig:profilo9284}.}\label{tab:parameters} \centering
\begin{tabular}{cc|}
  \multicolumn{2}{c}{fitness $f=2.20\pm0.05$}\\
  \hline\hline
  Parameter & Component 1 \\
  \hline
  $B$ (G) &  $164\pm88$ \\
  $\gamma$ $(^\circ)$ & $121\pm20$ \\
  $\chi$ $(^\circ)$ & $11\pm51$ \\
  $v_{LOS}$ (km/s) & $3.0\pm0.1$ \\
  $\alpha$ & 1.00 \\
  \hline
\end{tabular}
\centering
\begin{tabular}{ccc}
  \multicolumn{3}{c}{fitness $f=1.72\pm0.07$}\\
  \hline\hline
  Parameter & Component 1 & Component 2 \\
  \hline
  $B$ (G) &  $141\pm82^{\mathrm{a}}$ & $141\pm82^{\mathrm{a}}$ \\
  $\gamma$ $(^\circ)$ & $82\pm5$ & $40\pm26$ \\
  $\chi$ $(^\circ)$ & $36\pm5$ & $38\pm40$ \\
  $v_{LOS}$ (km/s) & $-52\pm1$ & $1.8\pm0.2$ \\
  $\alpha$ & $0.35\pm0.05$ & $0.65\pm0.05$ \\
  \hline
\end{tabular}
\centering
\begin{tabular}{cccc}
  \multicolumn{4}{c}{fitness $f=2.35\pm0.06$}\\
  \hline\hline
  Parameter & Component 1 & Component 2 & Component 3 \\
  \hline
  $B$ (G) &  $135\pm45^{\mathrm{a}}$ & $135\pm45^{\mathrm{a}}$ & $135\pm45^{\mathrm{a}}$ \\
  $\gamma$ $(^\circ)$ & $83\pm5$ & $124\pm19$ & $27\pm19$ \\
  $\chi$ $(^\circ)$ & $-10\pm2$ & $-16\pm12$ & $4\pm7$ \\
  $v_{LOS}$ (km/s) & $-53\pm1$ & $-30\pm2$ & $1.9\pm0.2$ \\
  $\alpha$ & $0.32\pm0.04$ & $0.21\pm0.04$ & $0.47\pm0.03$ \\
  \hline
\end{tabular}
\centering
\begin{tabular}{ccccc}
  \multicolumn{5}{c}{fitness $f=1.90\pm0.08$}\\
  \hline\hline
  Parameter & Component 1 & Component 2 & Component 3 & Component 4\\
  \hline
  $B$ (G) &  $106\pm46^{\mathrm{a}}$ & $106\pm46^{\mathrm{a}}$ & $106\pm46^{\mathrm{a}}$ & $106\pm46^{\mathrm{a}}$ \\
  $\gamma$ $(^\circ)$ & $80\pm7$ & $91\pm5$ & $103\pm5$ & $160\pm18$ \\
  $\chi$ $(^\circ)$ & $-17\pm30$ & $-17\pm30$ & $1\pm19$ & $35\pm11$ \\
  $v_{LOS}$ (km/s) & $-52\pm2$ & $-18\pm1$ & $1\pm2$ & $19\pm1$ \\
  $\alpha$ & $0.17\pm0.04$ & $0.23\pm0.05$ & $0.36\pm0.04$ & $0.24\pm0.04$ \\
 \hline
\end{tabular}
\centering
\begin{tabular}{cccccc}
  \multicolumn{6}{c}{fitness $f=1.79\pm0.11$}\\
  \hline\hline
  Parameter & Component 1 & Component 2 & Component 3 & Component 4 & Component 5 \\
  \hline
  $B$ (G) &  $244\pm55^{\mathrm{a}}$ & $244\pm55^{\mathrm{a}}$ & $244\pm55^{\mathrm{a}}$ & $244\pm55^{\mathrm{a}}$ & $244\pm55^{\mathrm{a}}$ \\
  $\gamma$ $(^\circ)$ & $11\pm11$ & $126\pm10$ & $89\pm6$ & $154\pm17$ & $154\pm18$ \\
  $\chi$ $(^\circ)$ & $15\pm47$ & $15\pm40$ & $12\pm40$ & $-19\pm9$ & $3\pm65$ \\
  $v_{LOS}$ (km/s) & $-25\pm1$ & $3.6\pm0.5$ & $25\pm1$ & $67.3\pm0.6$ & $100\pm1$ \\
  $\alpha$ & $0.08\pm0.01$ & $0.28\pm0.04$ & $0.23\pm0.04$ & $0.29\pm0.03$ & $0.11\pm0.03$ \\
  \hline
\end{tabular}
\begin{list}{}{}
\item[$^{\mathrm{a}}$] The $B$ values in all components were
    forced to be equal (in each table).
\end{list}
\end{table*}

\section{Discussion and conclusions}

We focus on the analysis and the inversion of Stokes profiles of the
\ion{He}{i} 10830~{\AA} triplet observed in an active region
filament during its phase of activity. We showed that even the most
complex observed profiles can be well reproduced by multi-component
atmospheres. Although we cannot completely exclude that the profiles
could be simply strongly turbulently broadened, the fits based on
multiple components are of clearly better quality (see below). We
tested the response of the numerical code H{\footnotesize
E}LI{\footnotesize X} for the inversion of Stokes profiles to work
with a high number of atmospheric components and free parameters.
This was necessary to fit not only the different observed He
components, but also the photospheric and telluric lines in the
wavelength range. We were able to retrieve the atmospheric
parameters for the individual He components at different levels of
accuracy. The values of the magnetic field strength, $B$, retrieved
from the fit presented here are in the range $100-250$~G. These
values are lower than those found by \citet{kuckein} for another
active prominence. This difference could be caused by the different
methods employed, because unlike \citet{kuckein} we inverted our
profiles including a scattering polarization correction (see
Sec.~\ref{sec:inversion}) and the magnetic field strength we
retrieved is a result of a coupling between many He components. The
difference could also be merely a reflection of the different
strengths of the two filaments' magnetic fields, or the analyzed
profiles possibly do not sample the highest field strength region in
the observed filament. The $v_{LOS}$ is well retrieved for all
profiles with a small error. The inclination angle $\gamma$ is
retrieved with a somewhat larger error, but the inclinations of the
different components of the magnetic field can often be
distinguished. The error bars on the azimuth angle, $\chi$, are
instead quite large, and it is often impossible to distinguish
between the azimuths of the different magnetic components. This is
mainly because of the complexity and the noise in the $Q$ and $U$
Stokes profiles. The information in the observations for retrieving
the azimuth is limited. We therefore refrain from conclusions about
the azimuthal direction of the magnetic field in the filament. There
is also insufficient information in the profiles to distinguish
between the field strength $B$ of the different components. Note
that because Stokes $V$ is generally much stronger than $Q$ and $U$,
$B\cos\gamma$ is probably the most reliably
determined quantity, although we do not explicitly tabulate it. \\
The analysis of the observed polarization of the \ion{He}{i}
10830~{\AA} multiplet in the filament, carried out by inverting the
Stokes profiles, reveals different unresolved
atmospheric components of the He lines, coexisting within each
spatial resolution element ($\sim1$~arcsec). The components are
distinguished by their Doppler shifts, which
generally differ by more than the sound speed. For more complex
analyzed profiles we also tried to find a solution considering a
broadened profile with some emission contributions distributed at
the appropriate places, but the fits to the observed profiles were
not as good as the ones we present in the paper. The
main problem is reproducing the $I$ and $V$ profiles at the same time. \\
In this active filament we find profiles requiring up to five atmospheric
components for a reasonable fit. Multiple unresolved magnetic components are
found not only in filaments, but also above pores and elsewhere in
active regions \citep{aznar,lagg1}, but so far in other data sets
never more than three components were needed. As \citet{lagg1} pointed
out, the geometrical interpretation of these multiple downflow
components is not clear-cut in the case of a line such as
\ion{He}{i} 10830~{\AA}, which is generally optically thin. Thus the
different components could be located at different horizontal
positions within the resolution element, implying considerable fine
structure in the filament. This fine structure in
filaments at a sub-arc s scale has been reported earlier. See, e.
g., \citet{tandberg} or \citet{heinzel}, for reviews. These fine
structures show a random motion with velocities of 5-10~km/s, as
deduced mainly from observations in the \ion{H}{$\alpha$} line. \\
An alternative explanation is that the various components are
located at different heights along the LOS. It may be argued that
the \ion{He}{i} line gets very strong in the filament and can
hardly be considered to be optically thin, so that one should not be able
to see through different layers so easily. We note, however, that
the line becomes optically thick only at wavelengths at which the
absorption is large, whereas it remains
optically thin at neighboring wavelengths. Consequently, if the different layers absorb at
wavelengths that are shifted by more than a Doppler width relative
to each other, this explanation remains possible even if individual
layers produce optically thick absorption. \\
Because in many cases the profiles have some optical thickness (they
are formed over a range of heights), it is possible that gradients
in velocity along the LOS can affect their shape. The influence of
these gradients cannot be treated in the simple model employed here. A
more general treatment with gradients might lead to good fits with a
reduced number of components. \\
The inversions support the idea that the He components correspond to
 plasma trapped in different magnetic field lines, which may well be
pointing in different directions along the LOS. Indeed, the
coupling of the magnetic field vector between the individual
components results in worse fits to the observed profiles, which
supports this interpretation. Note though that
some He profiles do not show a significant signal in Stokes $V$.
They describe intensity features without a polarization
counterpart.\\
Multiple magnetic components are fairly common in most parts of the
observed filament and are often associated with strong blue- or redshifts
corresponding to supersonic velocities. We measured downflow
velocities of up to 100~km/s and upflows of up to 60~km/s along the
line-of-sight. These supersonic up- and downflows always coexist
with a He atmospheric component almost at rest
($-10<v_{LOS}<+10$~km/s) within the same resolution element.
Sometimes strong blue- and redshifted components are found to
coexist in the same profile. Velocities of filament fine structures
reported in the literature are low compared with these observed
supersonic down- and upflows. \\
Strong motions and velocities in active region filaments were
previously observed in the He I 10830~{\AA} lines. Observations of
an erupting active region filament presented by \citet{penn1}
revealed the He I absorption line blue-shifted to velocities between
200 and 300~km/s. In the same observation, the filament also showed
internal motions with multiple Doppler components shifted by
$\pm25$~km/s. Redshifts of 30-60~km/s were reported by \citet{penn}
in an active filament during the decay phase of a flare in the He I
10830 absorption lines. We found the highest redshifts so far
measured with the \ion{He}{i} 10830~{\AA} line. In this respect, we
emphasize the importance of the TIP-II instrument, which was used for
the first time during this observation campaign. We were able to
observe the full range of He absorption signatures and measure the
highest downflow velocities thanks to its wider range of wavelengths
compared to the original TIP instrument.

\begin{acknowledgements}
We thank Luca Teriaca for helpful discussions. We also
thank Regina Aznar Cuadrado and Manolo Collados for their help
during the observing run at the German Vacuum Tower Telescope at the
Spanish observatory in Tenerife. CS thanks the IMPRS on Physical
Processes in the Solar System and Beyond for the opportunity to
carry out the research presented in this paper. This work was
supported by the program ``Acciones Integradas Hispano-Alemanas'' of
the German Academic Exchange Service (DAAD project number
D/04/39952), by the WCU grant No. R31-10016 from the Korean Ministry
of Education, Science and Technology, and by the ASI/INAF contract
I/05/07/0 for the program ``Studi Esplorazione Sistema Solare''.
\end{acknowledgements}

\bibliographystyle{aa}

\end{document}